# Cryogenic Compact Low-Power 60GHz Amplifier for Spin Qubit Control in Monolithic Silicon Quantum Processors

Michele Spasaro[1], Shai Bonen[2], Gregory Cooke[2], Thomas Jager[2], Tan D. Nhut[1], Dario Sufrà[1], Sorin P. Voinigescu[2], Domenico Zito[1]

[1]Dept. of Electrical and Computer Engineering, Aarhus University, Denmark
[2]Edward S. Rogers Sr. Dept. of Electrical and Computer Engineering, University of Toronto, ON, Canada
domenico.zito@ece.au.dk

*Abstract*—This paper reports the design and experimental characterization of a cryogenic compact low-power 60GHz amplifier for control of electron/hole spin qubits, as elementary building block for monolithic Si quantum processors. Tested at 2 K, the amplifier exhibits $S_{21}$ of 15 dB at 59 GHz, $BW_{3dB}$ of 52.5-67.5 GHz, and power consumption of 2.16 mW. Owing to the topology with inductorless active network, the amplifier has a compact core area of 0.18×0.19 mm$^2$.

## I. INTRODUCTION

Today's quantum processors (QPs) feature from tens [1] to 127 [2] physical qubits and support quantum algorithms with no more than a few thousand gates, i.e., quantum operations, limited by the qubit coherence time [3]. Quantum error correction [4] could overcome the limitations set by decoherence and pave the way to fault-tolerant quantum computers (QCs) [3, 4]. However, with current qubit technologies, a fault-tolerant QC would require about 1000 physical qubits per logical qubit with error correction and one million physical qubits [3, 4] to deliver the envisaged disruption of universal quantum computing in applications such as drug discovery [5] and cryptography [6].

Owing to the long coherence time [7] and the potential of CMOS technology to integrate a dense array of qubits in a single silicon (Si) die, electron/hole spin qubits in Si quantum dots (QDs) are a promising route towards highly scalable QPs [8, 9]. CMOS technology has also the potential for the monolithic integration of qubits and control and readout circuits, which is a promising solution to the wiring challenge [10].

Qubits are typically operated at cryogenic temperatures (CTs) lower than 0.1 K, in a cryostat with a cooling power that is largely insufficient with respect to the power dissipation of the control/readout circuits. For this reason, qubits operating at a few Kelvins [8, 11, 12], are key enablers for the practical implementation of monolithic quantum processors [4].

Since the higher thermal energy, $k_BT$, affects the coherence time, spin qubits operating at higher temperature should exhibit a larger Zeeman energy separation $(E_z)$ [8]. A train of pulses at the Larmor frequency $f_L = E_z/h$, where $h$ is the Planck's constant, allows the manipulation of the qubit spin state. For example, $E_z \approx 0.25$ meV corresponding to an $f_L$ of about 60 GHz [8].

This paper presents a cryogenic compact low-power 60GHz amplifier (AMP) in 22nm fully-depleted silicon-on-insulator (FDSOI) CMOS technology, as a building block for qubit control in monolithic QPs. The paper is organized as follows. Section II summarizes the control of Si spin qubits. Section III reports the main features of the amplifier. Section IV reports the experimental characterization at room temperature (RT) and 2 K. Section V draws the conclusions.

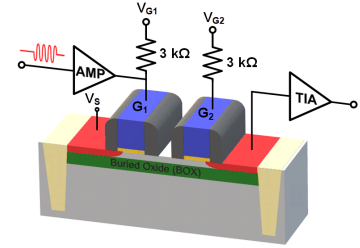

Fig. 1. Example of integrated spin qubit based on a double quantum dot with control and readout circuits in FDSOI CMOS technology.

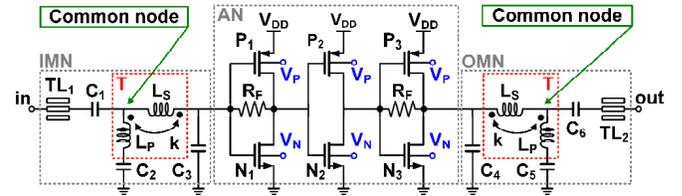

Fig. 2. AMP circuit with the active network (AN), and input matching network (IMN) and output matching network (OMN) with spiral transformers (T).

## II. SI SPIN QUBIT CONTROL

Fig. 1 shows an example of integrated spin qubit with control and readout circuits in FDSOI CMOS technology [8, 13], comprising a double quantum dot (DQD), an amplifier (AMP), and a readout transimpedance amplifier (TIA). The DQD hosts the electron/hole spin qubit, whose spin state is manipulated through electric-dipole spin resonance (EDSR) [13]. AMP serves as a driver to deliver the 60GHz spin-manipulation signal generated off-chip to the control gate ($G_1$) of the DQD, providing a 50Ω input impedance and delivering an output voltage of a few mV [14] at the gate $G_1$. As the integration of the AMP in monolithic QPs is the ultimate target, AMP should be low power and compact in size, with acceptable noise and bandwidth. The AMP should exhibit an $S_{21}$ of 10 dB, in order to compensate for the conversion loss of the off-chip up-converter (UC). The study in [9] on 10nm FDSOI hole-spin qubits predicts the feasibility of single-qubit gates with $f_L = 60$ GHz, Rabi frequency $f_R = 60$ MHz, and zero-to-peak gate voltage $V_g = 10$ mV. Following the analysis in [14], to limit the infidelity contribution of the noise added by UC and AMP to $125\times10^{-6}$ for an overall fidelity of 99.9%, the AMP should exhibit NF < 12 dB. For $\pi/2$-rotation gates, the BW must exceed $8\times f_R = 480$ MHz. A 10GHz BW allows a wide flexibility on $f_L$.

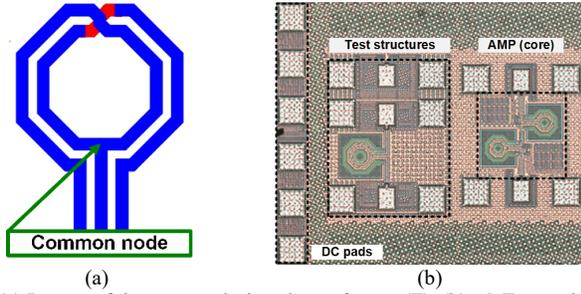

Fig. 3. (a) Layout of the custom-designed transformer (T). (b) AMP test-chip together with test structures for verification and calibration.

### III. AMPLIFIER FEATURES

Fig. 2 shows the AMP circuit designed in the commercially available 22nm FDSOI CMOS technology. The AMP comprises a broadband inductorless active network (AN), and feedforward matching networks (MNs) at the input (IMN) and output (OMN). The MNs are designed for maximum power transfer to 50Ω impedance, and the IMN is designed for minimum cascaded noise figure [15]. The MNs allow us to carry out the test of the AMP as standalone circuit, but may be not needed in the real application as qubit driver in monolithic QPs; also, the number of active stages could be reduced, leading to further area and power saving. All devices are super-low $V_t$ (slvt) MOSFETs with total gate width of 3.9 μm. As the AN is inductorless, and the spiral inductors of the MNs are coupled as in an integrated transformer, the AMP is very compact in size. Fig. 3(a) depicts the layout of the custom-designed transformer.

Because the design kit does not simulate at 2 K and the foundry models are only guaranteed above 225 K, the design was carried out at RT for validation, targeting a gain of about 10 dB, and expecting better performance at CT [8, 16], owing to the increase in Q-factor of the T [17] and $f_{max}$ of the transistors [16], provided that the bias voltages are adjusted to get the same current density $J$ as at RT. At RT, post-layout simulation (PLS), including 3D EM simulation of the T, shows that the AMP exhibits a gain of 10.4 dB at 56.9 GHz with a $J$ of 0.21 mA/μm, i.e., between the optimum-$NF_{min}$ $J$ of 0.15 mA/μm and peak-$f_{max}$ $J$ of 0.25 mA/μm, i.e., temperature-invariant [16], and a supply voltage ($V_{DD}$) of 0.88 V. The power consumption ($P_{DC}$) of 2.16 mW. To take advantage of the forward body bias (FBB) capability of slvt MOSFETs and set the drain-source voltages $V_{DS} = V_{DD}/2$, the back-gate voltages amount to: $V_N = 0.5$ V, $V_P$ = -0.6 V. Lowering $V_{DD}$ to 0.8 V leads to the $J$ of 0.15 mA/μm, a gain of 8.3 dB at 57.1 GHz, with a $P_{DC}$ of 1.40 mW. Since $|V_t|$ increases as the temperature decreases [16], the $J$ required at 2 K is obtained by adjusting $V_N$ and $V_P$, as reported hereinafter.

### IV. EXPERIMENTAL CHARACTERIZATION

Fig. 3(b) reports the chip micrograph. AMP has a core area of $0.18\times0.19$ mm$^2$. On-chip experimental characterization has been carried out at RT and CT, in order to validate the design carried out at RT and measure the performance at 2 K.

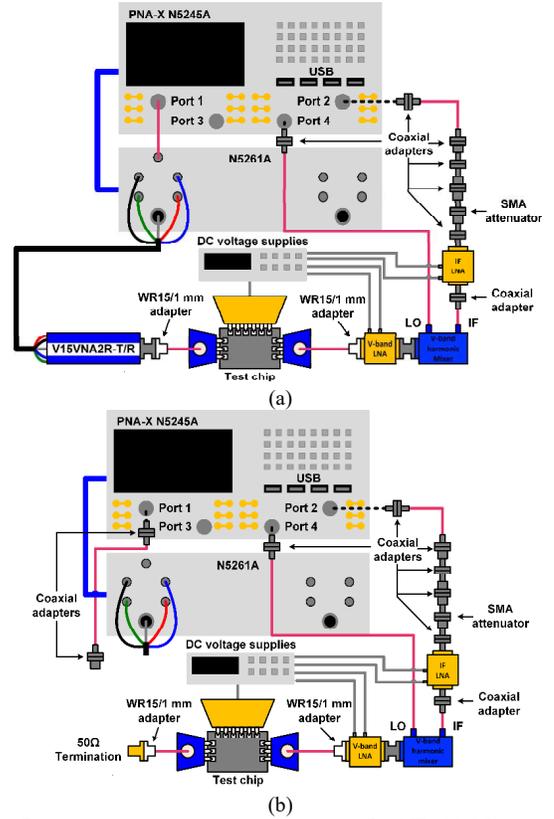

Fig. 4. Room-temperature measurement setup for NF. (a) Measurement of conversion gain. (b) Measurement of output noise power density.

#### A. Room Temperature

The RT measurements have been performed with the Keysight PNA-X N5245A and mm-head controller N5261A, two OML mm-wave modules V15VNA2R-T/R, for banded measurements in V-band, and FormFactor Infinity Probes GSG i110. Power calibrations were carried out with the power meter N1914A, 50MHz-50GHz power sensor N8487A, and the V-band power sensor V8486A by Keysight Technologies.

The noise figure (NF) of AMP has been measured by means of the mm-wave cold-source NF measurement methodology [18]. Fig. 4(a) reports the measurement setup for the conversion gain of AMP plus the downconverter (DC), which comprises the Quinstar V-band low noise amplifier (LNA) QLW-50754530-12, the Keysight V-band harmonic mixer 11970V, the Planar Monolithic Industries LNA PEC-42-1G40G-20-12-292MM, and a 12dB SMA attenuator. Fig. 4(b) reports the setup for the measurement of the output noise power density (ONPD) with the input-port of AMP connected to a WR15 50Ω termination. The LNAs in the DC are essential for the ONPD to exceed the noise floor, while the SMA attenuator prevents the overload of the low-noise receiver of the PNA-X. Unlike from [18], the measurement setup in Fig. 4(b) uses a two-port mm-head controller and the PNA-X was configured in standard PNA-X mixer-mode for the measurement of ONPD in a "Noise Figure Converter" channel, so overcoming the need to edit a PNA-X registry. In Fig. 4(b), an RF cable is connected to Port 1, for calibration purposes. As required by the procedure in [18], we have also characterized the DC. The conversion gain of DC is measured with the setup as in Fig. 4(a), where the test chip is replaced by the through-line (thru) in the impedance standard substrate. The ONPD of DC is measured as in Fig. 4(b), but with the 50Ω termination connected at the input port of the V-band LNA. Fig. 5 reports the measurement results for $V_{DD} =$ 0.88 V. Fig. 5(a) shows the measured S-parameters and NF, as well as NF resulting from PLS. AMP exhibits a center frequency $f_0$ of 57.25 GHz, $S_{21}$ of 9.9 dB and a minimum NF of 7.1 dB in its 3dB bandwidth. Fig. 5(b) reports the measured transducer gain ($G_T$) as a function of the input power ($P_{in}$) at $f_0$. The AMP exhibits an input-referred $P_{1dB}$ ($IP_{1dB}$) of -21.8 dBm. For $V_{DD} = 0.8$ V, $f_0 = 57.25$ GHz, $S_{21} = 7.3$ dB and $IP_{1dB} = -20.2$ dBm.

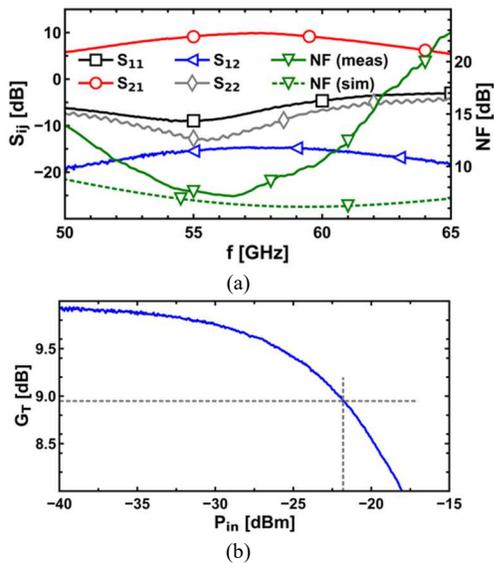

Fig. 5. Room-temperature performances of AMP for $V_{DD} = 0.88$ V: (a) S-parameters (measured) and NF (measured and simulated). (b) Measured transducer gain $G_T$ vs. input power ($P_{in}$) and $P_{1dB}$ at center frequency $f_0$.

### B. Cryogenic Temperature

Fig. 6 shows the cryogenic measurement setup from DC to 70 GHz with a Lake Shore Cryotronics CPX-VF-LT probestation. The adjusted voltages are obtained from the DC transfer characteristics of the MOSFETs measured at 300 K and 10 K, and considering that the characteristics of MOSFETs in saturation do not change significantly as the temperature is lowered from 10 K to 2 K [8, 16]. For $J$ equal to 0.21 mA/μm with $V_{DD} = 0.88$ V at 2 K, the required back-gate voltages are $V_N = 998$ mV and $V_P = -827$ mV; for $J = 0.15$ mA/μm and $V_{DD} = 0.8$ V at 2 K, the required back-gate voltages are $V_N = 1.20$ V and $V_P = -1.01$ V, leading to same power consumption as at RT.

Fig. 7(a) and Fig. 7(b) show the S-parameter measurements at 2 K for 0.88 V and 0.8 V, respectively. The PNA-X Port 1 output power was set to secure on-chip calibration and equal power level on the input pad within the linear operating region of the AMP. With $V_{DD} = 0.88$ V, the AMP exhibits an $f_0$ of 59 GHz, $S_{21}$ of 15 dB, and 3dB band from 52.5 to 67.5 GHz. With $V_{DD} = 0.8$ V, the AMP exhibits an $f_0$ of 58 GHz, $S_{21}$ of 13.4 dB and a 3dB band from 52.3 to 64.7 GHz. As expected, the slight positive frequency shift with respect to the RT is the effect of the Q-factor increase and the self-inductance reduction [17].

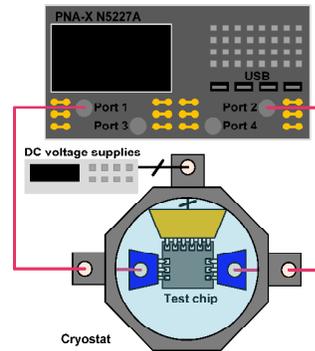

Fig. 6. Closed-cycle helium cryogenic on-chip measurement setup from DC to 70 GHz, including multi-contact DC and GSG RF probes.

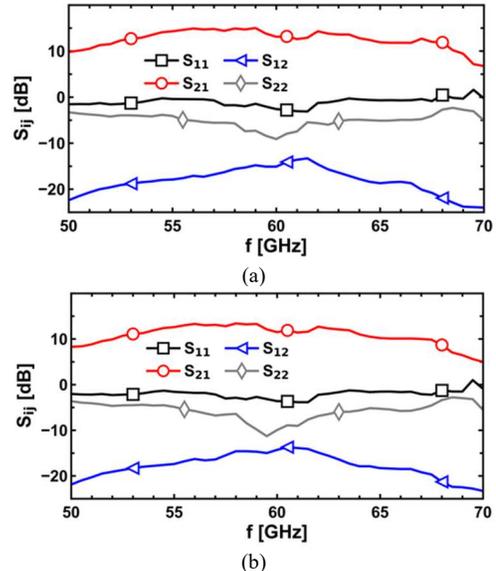

Fig. 7. Measured S-parameters at 2 K: (a) $V_{DD} = 0.88$ V and (b) $V_{DD} = 0.8$ V.

As the equivalent noise temperature of passive networks is proportional to the physical temperature [15], which reduces by two orders of magnitude from RT to CT, and the $T_{Nmin}$ of MOSFETs typically decreases by one order of magnitude [19], we expect the equivalent noise temperature of AMP to decrease by at least one order of magnitude from RT to CT.

### V. CONCLUSIONS

A compact low-power 60GHz amplifier for monolithic QPs has been designed in 22nm FDSOI CMOS and characterized experimentally at 300 K and 2 K. Compact area and low-power consumption are promising for qubit control ICs at 60 GHz. The experimental measurements validate the design methodology at room and cryogenic temperatures, and confirm the expected gain improvement at cryogenic temperatures.

### ACKNOWLEDGMENTS

The authors are grateful to Keysight Technologies for their support through the donation of equipment and cad tools; Dr. C. Kretzschmar, Dr. P. Lengo, Dr. B. Chen (GlobalFoundries) for the technology support. This work was co-funded by the European Commission through the European H2020 FET Open project IQubits (G.A. N. 829005).